\documentclass[aps,prl,twocolumn,citeautoscript,superscriptaddress]{revtex4}
\usepackage{amsmath}
\usepackage{amssymb}
\usepackage{graphicx,bm}
\usepackage[caption=false]{subfig}
\usepackage{verbatim}
\usepackage{epstopdf}
\usepackage{dcolumn}
\usepackage{floatrow}
\usepackage{float}
\usepackage{booktabs}
\usepackage[utf8]{inputenc}
\usepackage{array}
\usepackage{makecell}
\usepackage{booktabs}
\usepackage{multirow}
\usepackage{dcolumn}
\usepackage{footnote}

\usepackage{lipsum}

\usepackage{xcolor}
\usepackage{soul}
\usepackage{hyperref}
\hypersetup{
	colorlinks=true,
	linkcolor=magenta,
	filecolor=violet,     
	urlcolor=blue,
	citecolor=red
}
\UseRawInputEncoding

\begin{document}

\preprint{APS}

\title{Development of a cylindrical mirror analyzer electron spectrometer and associated data acquisition system to study inner shell electron emission following ion-atom collision}

\author{Rohit Tyagi}
%\email{rtyagi@iitk.ac.in}
\affiliation{Department of Physics, Indian Institute of Technology Kanpur, Kanpur - 208016, India}
\author{A. H. Kelkar}
\email{akelkar@iitk.ac.in}
\affiliation{Department of Physics, Indian Institute of Technology Kanpur, Kanpur - 208016, India}


\date{\today}% It is always \today, today,

\begin{abstract} In this paper we report on the development and performance of a cylindrical mirror analyser electron spectrometer for ion atom collision experiments. A low cost data acquisition system using Arduino microcontroller has also been developed and tested. We have measured the Auger emission spectra for various gaseous targets in collision with 1 MeV proton beams. Relative total Auger emission cross sections have also been measured for N$_2$ molecular target as a function of proton energy.
\end{abstract}

\maketitle

%\tableofcontents

\section{Introduction}
Electron Spectroscopy involves the study of the energy and angular emission pattern of secondary electrons emitted from atoms, molecules, clusters and solid surfaces upon excitation via charged particles (electrons or heavy ions) and photons. In ion-atom collision, the nature of Coulomb interactions manifests in various ionization processes such as soft electron emission, binary encounter and Auger electron emission etc. Study of these process is useful in atomic and molecular physics, material science, space research, and plasma diagnostics etc. In the last couple of decades, there has been an increase in interest in probing state-selective atomic and molecular phenomena such as Inter-atomic/Intra-molecular Coulombic decay (ICD) and molecular ion dissociation dynamics \cite{Murphy1988} using kinematically complete ion-electron yield measurements \cite{JUllrich_1997}. To study these processes, an electron energy analyzer is usually combined with a recoil ion momentum spectrometer\cite{JUllrich_1997,Albert2005i,Saha2013}.Various designs for an electron spectrometer, for example, hemispherical analyzer, parallel plate analyzer, cylindrical analyzer, magnetic bottle electron spectrometer etc are routinely employed in electron spectroscopy investigations. 

Here we present the development of a new electron spectrometer, based on a cylindrical mirror analyzer, for measuring secondary electron emission from gaseous targets induced by fast ion impact. A cylindrical mirror electron analyzer (CMA) has a simple design, excellent sensitivity, and low signal-to-noise ratio. The set-up consists of an ultra high vacuum compatible cylindrical mirror analyzer (CMA), designed and assembled at the 1.7 MV Tandetron accelerator laboratory at IIT Kanpur, to measure electron yields as a function of electron energy for fast ions interacting with atoms/molecules. In the following sections we will discuss the design parameters, simulation and performance of the CMA in detail. 
%The new spectroscopy set-up has been tested in a high-vacuum RIMS (Recoild ion momentum Spectroscopy) chamber that has been installed $20^0$ beam line in the TAL Laboratory.%
In addition to the CMA, we will also dicuss the implementation of an automated data acquisition and control system, based on Arduino Uno microcontroller. 
%Performance of the new analyzer, we have measured Auger electron emission from 1 MeV protons incident on atomic Ar, diatomic N$_2$, tri-atomic CO$_2$ and polyatomic CH$_4$ gas target and compared with well-known emission spectra. Ultimately, In the future, the new electron spectroscopy system will be augmented with the RIMS spectrometer to perform kinematically complete ionization/fragmentation studies.%
\\
%In this paper we will present design, simulation and working of a new electron energy analyzer which has been developed for measuring emission from gas targets induced by fast ion impact. This set-up consists of a high-vacuum compatible cylindrical mirror analyzer (CMA), designed and fabricated in the Tandetron Accelerator Laboratory (TAL) at IIT Kanpur. The CMA is made up of two coaxial cylinders in which outer cylinder has diameter 84 mm and radial distance in two cylinder is 20 mm. inner cylinder has two 3 mm slits separated by a distance of 77 mm. The new spectroscopy set-up has been tested in a high-vacuum RIMS chamber that has been installed in the 20$^0$ beam line of the accelerator. In addition to the new analyzer, data acquisition and experiment control system, based on Arduino uno and ESP-32 micro controllers has been developed. We have analyzed the performance of the spectrometer, measuring Auger electron emission from Argon, Nitrogen and CO2 gas targets in 1MeV proton collision. In future the new electron spectroscopy system will be augmented with the RIMS spectrometer to perform kinematically complete ionization/fragmentation studies.
\section{Experimental set-up and CMA design}
%\subsection{Experiment}
Performance characterization of the CMA was carried out by studying proton collision induced Auger electron emission for various atomic and molecular targets. The experiments were conducted at the 20$^o$ beam line of 1.7 MV tandetron accelerator facility \cite{HVEE} at IIT Kanpur. Intitally a beam of H$^-$ ions, produced using a duoplasmatron source, is accelerated to a potential of 10 kV - 15 kV. this beam is mass analyzed using a 90$^o$ magnet before entering the tandem accelerator. In the tandem accelerator the H$^-$ beam gets accelerated to the terminal potential ($TV = 200$ kV $- 1.7 $MV), where it collides with high pressure nitrogen gas. The collision process strips the H$^-$ ions of the electrons, leading to creation of H$^+$ beam which is further accelerated by the same potential in the outgoing channel. In general, for the tandem accelerator, the final energy of the ion beam is given as $(q+1)TV$, where $q$ is the ion charge state and $TV$ is the terminal voltage. The positive ion (proton) beam is further focused using an electrostatic quadrupole triplet and directed toward the 20$^o$ beam line using a switching magnet. The 20$^o$ beam line is equipped with a set of steering plates for alignment of the ion beam. In addition, the exit end of the beam line is slightly off centered from the beam line   and a set of deflector plates are used to deflect the ion beam, thereby filtering out the neutrals in the main ion beam created during its passage though the beam line. The filtered projectile ion beam is collimated to a size of 2$\times$2 mm$^2$ using a pair of four jaw slits before entering the main scattering chamber. A differential pumping chamber is connected between the four jaw slits to maintain high background vacuum ($< 5\times10^{-8}$ mbar) in the main scattering chamber. The CMA is mounted in the scattering chamber with its symmetry axis orthogonal to both, the ion beam and target gas jet as shown in fig \ref{fig1}. 
%The experiments have been performed in a 6 way cross stainless-steel scattering chamber. CMA was mounted as shown in fig., chamber is connected in the $20^0$ beam-line of 1.7MV Tandetron accelerator facility at IIT Kanpur. H- ion beam from duo plasmatron source has been extracted by applying 10kV extraction voltage and after mass selection using a $90^0$ magnet beam is injected to the accelerator which has potential gradient from ground to high voltage to ground at the end. in the middle H- ion gets stripped using N2 gas and charge exchange process provide H+ ion which subsequently gets accelerated from middle high potential to ground. This doubly accelerated beam is focused using quadruple electrostatic lens and steer to the $20^0$ beam line using a switching magnet. To separate neutral from the beam which could have formed due to charge exchange process beam is steered. The high vacuum 6 way cross stainless-steel chamber used for the experiments was specifically designed for measuring the scatter er from gas targets. A schematic 3D projected view  of the chamber is shown in Fig 1. Two-four jaw slits are positioned before the entrance of the chamber to achieve a collimated 2x2 mm beam for interacting target gas molecules. Proton beam after interacting with the target molecules travels through the chamber and get collected by a Faraday cup at the end of the chamber. Base pressure of the chamber was maintained to ~10$^{-8}$ mBar using Pfeiffer Vacuum turbo molecular pump.%
\subsection{Cylindrical mirror analyser}

\begin{figure*}
    \centering
    \includegraphics[width =\textwidth]{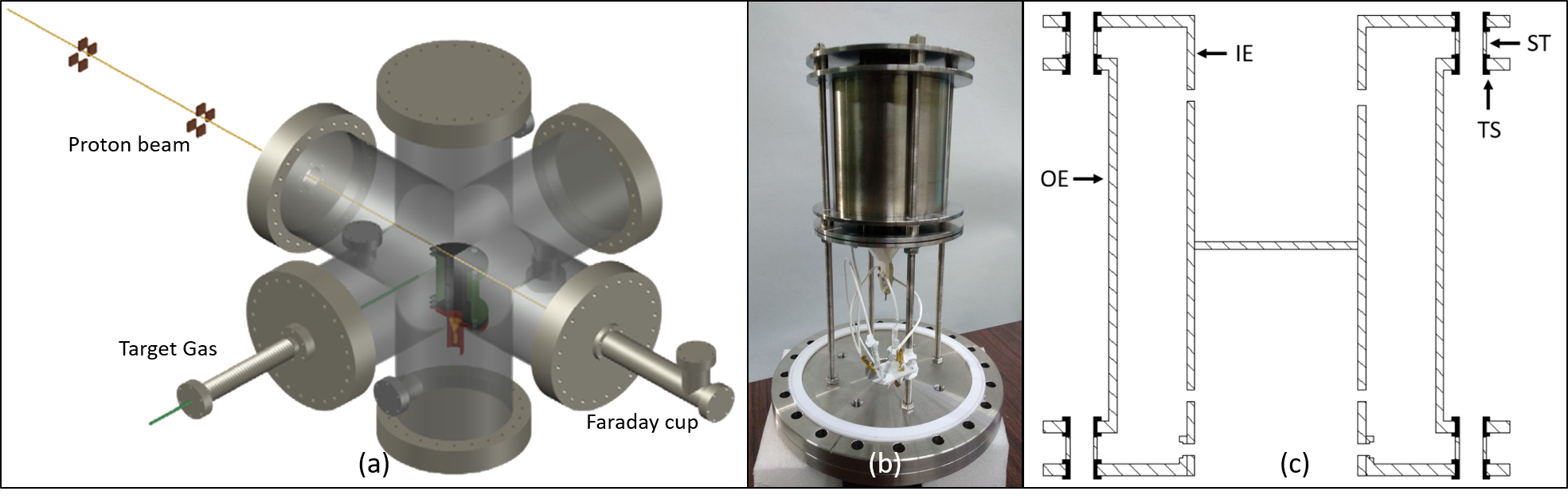}
    \caption{Schematic of experimental setup. In Left Schematic of the the CMA assembly. OE (outer electrode) cylinder diameter is 84 mm and length 96 mm. IE (Inner electrode) diameter is 40 mm. slit to slit distance is 77 mm. Total length of CMA is 118 mm. In middle Assembled CMA mounted on CF-150 flange is shown. In Right, CAD drawing of scattering chamber is shown.  }
    \label{fig1}
\end{figure*}

The Cylindrical mirror analyzer (CMA) (see fig \ref{fig1}) consists of two coaxial cylindrical shells made of non-magnetic stainless steel. A conical electron stopper is also connected to the inner cylinder to improve the energy resolution (discussed later).  CMAs are widely used as electron energy analyzers and their ideal design parameters are well known. Electrons emitted from the interaction zone, located on the axis of the CMA enter through a slit on the inner cylinder electrode, traverse the electric field region between the inner and outer cylinder and exit through another slit on the inner cylinder. The strength of electric field and distance between the two slits determine the pass energy of the electrons. The optimum location of the electron source (interaction region) on the axis is such that it subtends an entrance angle of 42.3$^o$ with respect to the axis of the cylinder \cite{Sar‐El1971}. However, due to geometrical constraints, the interaction zone in the present setup is located 30.5 mm away from first slit of the inner electrode. The entrance angle of the electrons, therefore, is 35.5$^o$ in our setup. The design parameters of the CMA have been chosen such that it can be coupled later with a recoil ion momentum spectrometer setup \cite{Duley2022} for performing combined recoil ion - electron spectroscopy following molecular fragmentation. 
%The cylindrical mirror analyzer (CMA) shares similarities with the parallel plate analyzer, differing in that its deflection plates take the form of coaxial cylinders. In fact, the parallel plate analyzer can be regarded as a specific case of the cylindrical mirror configuration. as shown in Figure 2., the typical geometry of the CMA places the interaction region (electron source) on the axis, where particles emitted within a cone defined by a polar angle traverse an annular slot in the inner cylinder. Particles with energy E experience deflection, ensuring their passage through an exit slot and subsequent focusing onto an image along the axis. The CMA exhibits double focusing, such that the image of a point at the source manifests as a point at the detector. An evident advantage of this analyzer lies in its ability to collect particles at any azimuthal angle. This feature enhances the versatility of the CMA in research applications.%
Using the geometrical parameters of the CMA and classical trajectory calculation for motion of a charged particle in cylindrical capacitor field \cite{Moore2009}, we can obtain the value of electrostatic potential on the outer electrode for transmission of electrons with energy $E_e$ (in eV) as: 
\begin{equation}
    \mathrm{V_{outer} =  \frac{E_e \ln\frac{R_2}{R_1}\sin^2\theta}{\ln(\frac{z}{4R_1}\tan\theta+1)}}
    \label{eq1}
\end{equation}
or, 
\begin{equation}
    \mathrm{V_{outer} = 0.48 E_e}
    \label{eq2}
\end{equation}
Here, $R_1$ and $R_2$ are the radii of inner and outer cylinders, respectively, $\mathrm{z}$ is the distance between the two slits and $\mathrm{\theta}$ is the entrance angle. 

In order to keep the interaction zone free of any electric field, the inner electrode was kept at ground potential. the outer surfaces of the inner electrode and electron stopper as well as the inner surface of the outer electrode were coated with carbon soot from an ethylene welding torch. the carbon soot coating helps in suppressing the emission of secondary electrons within the CMA due to collision of high energy electrons with the walls of the electrodes. A channel electron multiplier (CEM) detector is placed 24 mm away from the exit slit of inner cylinder (on the axis of CMA) to detect the electrons. 

%For axial extent of interaction region 'w' and L is separation between entry and exit slit the resolution of axial focusing CMA is given by
%\begin{equation}
 %   \frac{\Delta E}{E} = \frac{w}{L}
%\end{equation}
%For increase in the size of the interaction region slit width can be taken instead of w to calculate resolution.
\begin{figure}
    %\centering
    \includegraphics[width = \textwidth]{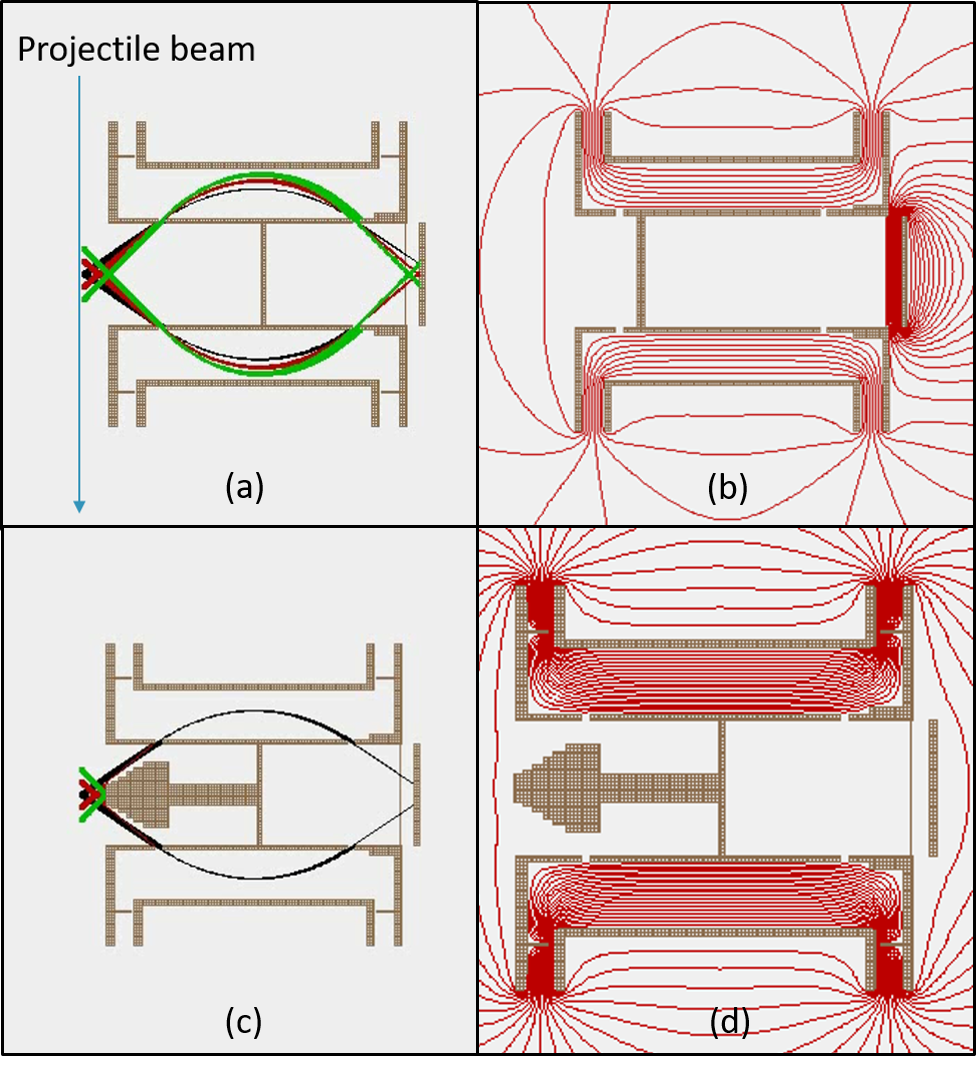}
    \caption{ Electron trajectories In (a) with stopper and (c) without stopper, in (b) and (d) potential contour lines are shown .}
    \label{fig2}
\end{figure}

\subsection{SIMION Simulation}

\begin{figure}
    \centering
    \includegraphics[width = \textwidth]{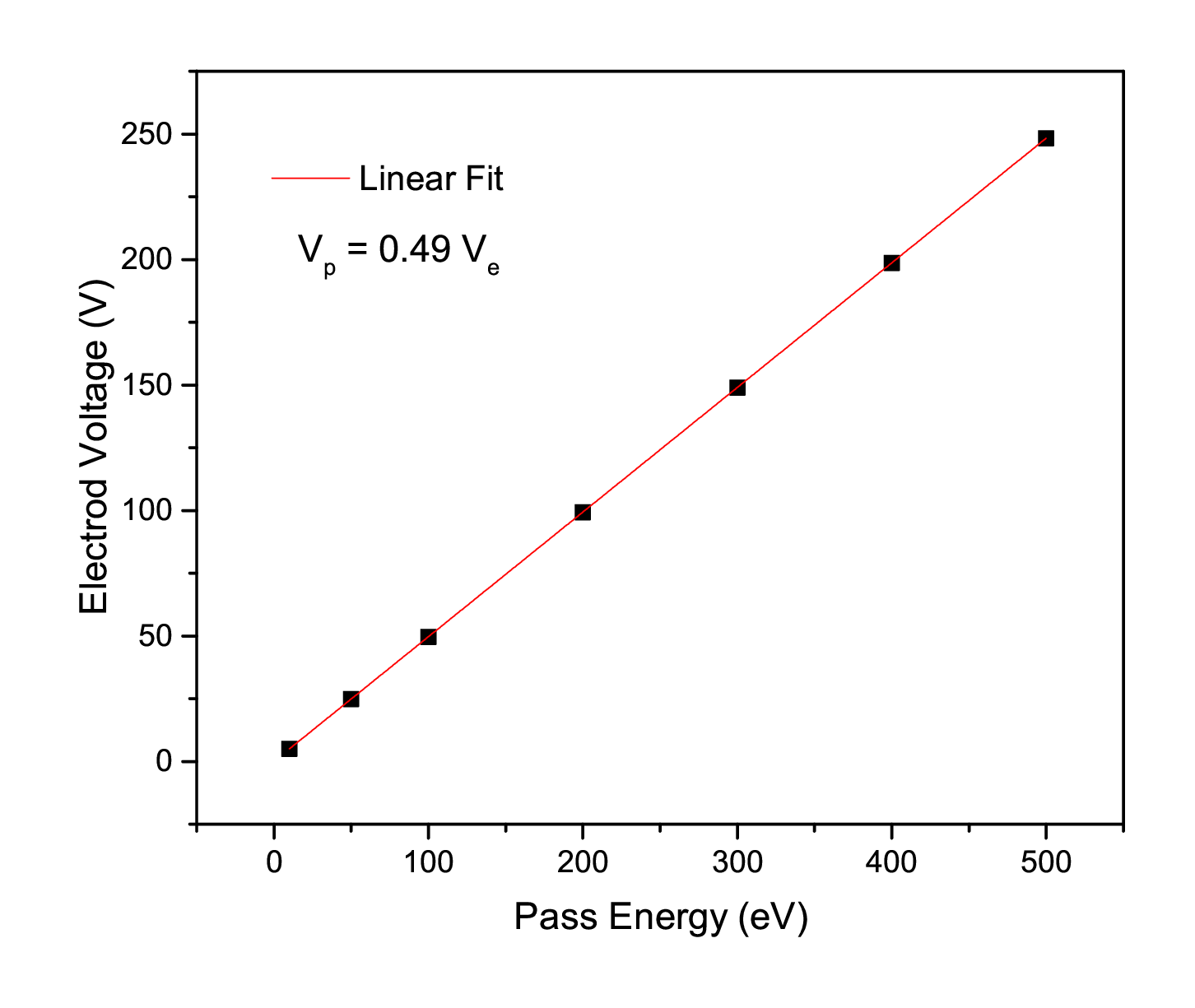}
    \caption{ Outer Electrode potential(negative) vs pass energy of electrons. Data is generated using SIMION sofware.}
    \label{fig3}
\end{figure}

The design of the CMA was optimized by performing particle trajectory simulations using SIMION 8.0 \cite{simion}. For electron trajectory simulation, the inner electrode was kept at ground potential and a negative potential was applied to the outer electrode. The interaction zone, kept 30.5 mm away from the entrance slit, was simulated as a spherical region of 2 mm radius. Thousands of electrons were generated in this region with a kinetic energy distribution {$\mathrm{E_e \pm 10\%}$. The particle velocities (direction) were constrained within an emission cone with half angle $\mathrm{\theta \pm 1^o}$, where $\theta$ is the geometric acceptance angle of the CMA. This approach aimed to streamline the simulation process and reduce computation time. 

In fig \ref{fig2} we have shown representative electron trajectories though the CMA. In the first design iteration (see fig \ref{fig2} (a) and (b)) it was observed that for an extended interaction zone, electrons of kinetic energies different from the pass energy of the CMA are also able to transmit though the CMA. this results in very poor energy resolution of the CMA. In the laboratory experiments, the interaction zone is defined as the overlap between the projectile ion beam and the effusive target gas jet, which may extend beyond 2 mm. The background gas density in the scattering chamber would also contribute as extended interaction zone along the path of the projectile ion beam. This poses a major challenge in achieving desired energy resolution of the analyzer. To alleviate this, we modified the design of the CMA by introducing a conical electron stopper (see fig \ref{fig2} (c and d)). the electron stopper was kept at the same potential as the inner electrode. As evident from the simulated trajectories (see fig \ref{fig2} c) the electron stopper creates a narrow annular region for electrons to enter the CMA, thereby restricting the acceptance of electrons emitted in a limited interaction volume on the axis of the spectrometer. Introduction of the electron stopper results in improved energy resolution of the CMA and from the simulated trajectories, we obtained the energy resolution of the CMA $\mathrm{\sim 3\%}$.

In fig \ref{fig3} we have plotted the voltage on the outer electrode vs the electron kinetic energy. The voltage varies linearly with the electron kinetic energy, as expected from equation \ref{eq1} above. A linear fit to the simulated data in fig \ref{fig3}, gives a slope of 0.49, which agrees well with equation \ref{eq2} for the present CMA geometry. 

\subsection{Data acquisition system}

\begin{figure*}
    \centering
    \includegraphics[width =\textwidth]{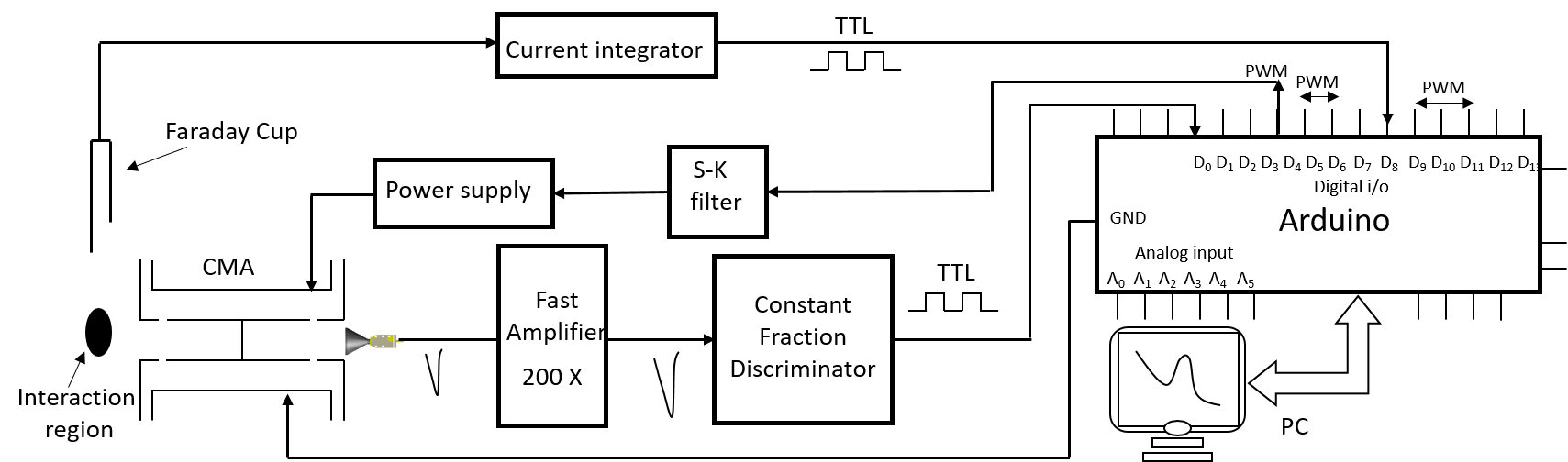}
    \caption{Schematic of Data Acquisition system.}
    \label{fig4}
\end{figure*}

\begin{figure}
    \centering
    \includegraphics[width =0.9\textwidth]{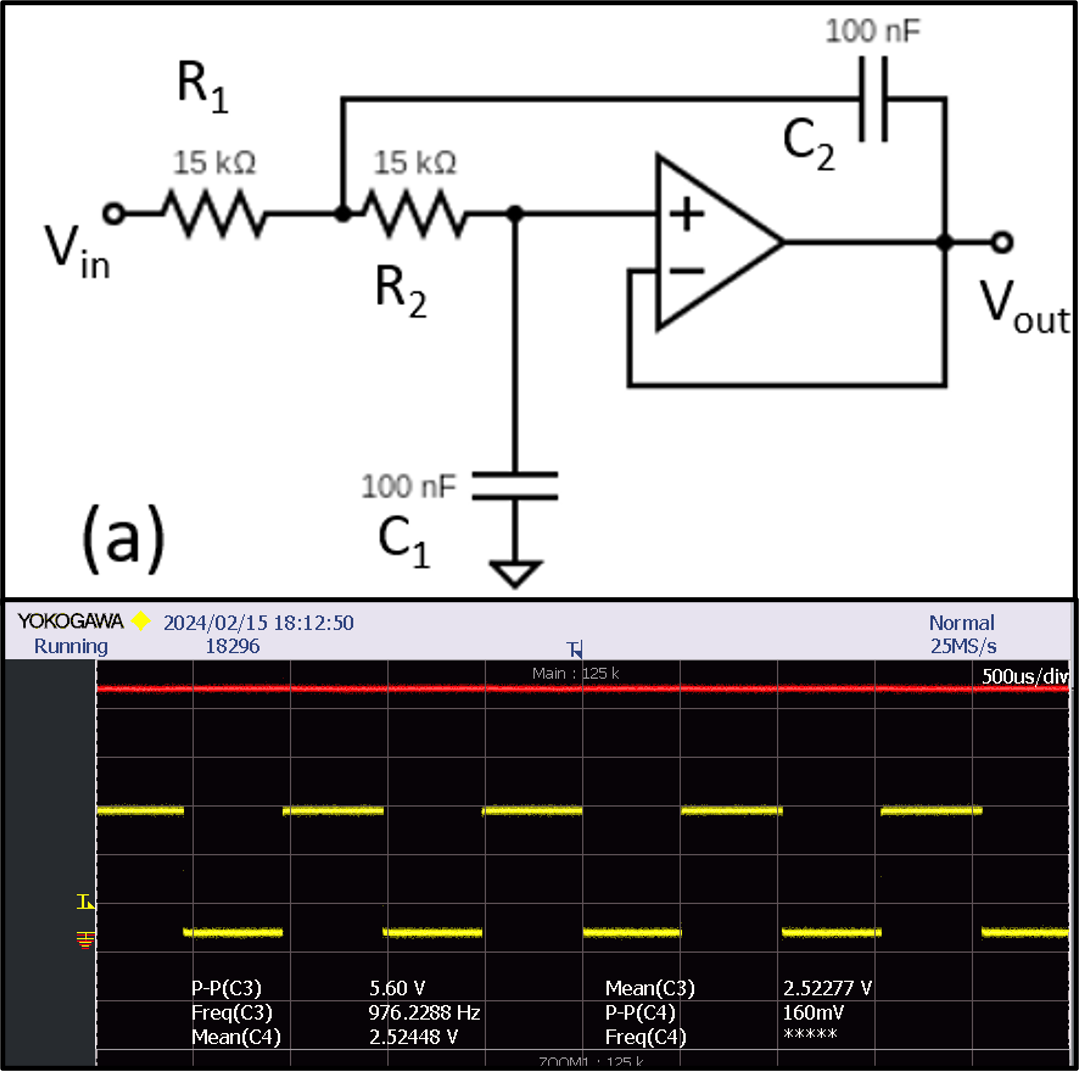}
    \caption{(a) Schematic of the circuit for Arduino PWM to Analog signal converter. (b) PWM output of Arduino with 50\% duty cycle (Yellow square wave signal) and analog dc output of the sallen-key filter (Red solid line).}
    \label{fig5}
\end{figure}

\begin{figure}
    
    \includegraphics[width =\textwidth]{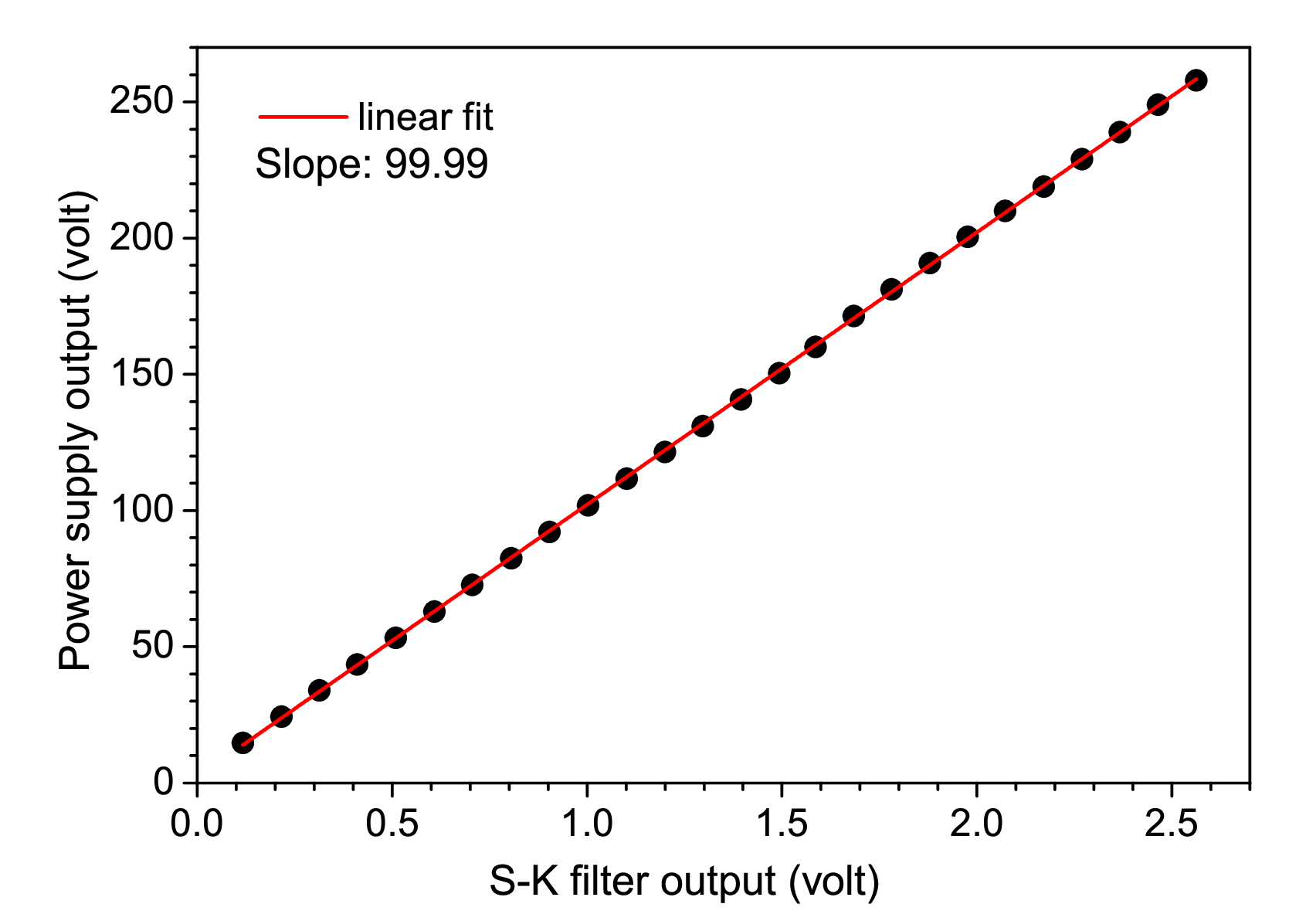}
    \caption{Sallen-Key filter output Vs power supply output. Black rectangular points show the measured values. Red solid line is a linear fit to the data. }
    \label{fig6}
\end{figure}

A schematic diagram of the data acquisition system is shown in figure \ref{fig4}. During the experiment, the outer electrode of the CMA is floated at the desired voltage using an externally controlled ortec power supply (). Electrons passing through both slits of the CMA are detected by the CEM detector, which generates a pulse of $\sim$ 10 mV ($\sim$ 20 ns) for each incident electron. The CEM is operated in saturation mode for maximum detection efficiency. The output signal of the CEM is amplified using a fast amplifier (Ortec ) and fed to a constant fraction discriminator (CFD). The CFD generates a corresponding TTL output which is stored in the PC after digitization. The projectile ion beam current falling on the Faraday cup is measured using a charge integrator module (ortec). The charge integrator also generates TTL pulses which are recorded in the PC after digitization. In order to record the electron spectrum over a given energy range, automatically, the data acquisition system is required to handle three signals. i) Generate an analog control signal for the dc power supply connected to the outer electrode of the CMA, ii) digitize and record the TTL signal corresponding to the CEM detector output and iii) record the TTL output from the charge integrator module. A low cost data acquisition system was designed in-house, to automate the process using off-the-shelf Arduino microcontroller \cite{Arduino} and Raspberry Pi \cite{RaspberryPi} personal computer.

The Arduino microcontroller is a versatile open-source platform designed for electronics projects. It features analog and digital input/output pins, enabling it to interface with sensors, collect data, and communicate with other devices. The programming interface for Arduino Uno is highly user-friendly extensive community support makes it a popular choice for creating customized data acquisition setups. Arduino Uno microcontroller has a digital i/o pins which can be used to record the TTL signals. It also has a 12 bit digital-to-analog converter and analog output pins which generate a corresponding PWM output. The analog remote signal for controlling the outer electrode power supply was generated using the PWM output of the Arduino Uno (Pin 12). The PWM (pulse width modulated) output is a 5 V square wave signal with varying duty cycle, corresponding to each analog value. We designed a second order Sallen-Key low-pass filter \cite{Sallen1955} convert the PWM output into a true analog dc signal. This signal is then used to control the ortec power supply connected to the outer electrode of the CMA. The Schematic circuit of the Sallen-Key filter is shown in figure \ref{fig5}a. This circuit incorporate a basic IC 741. The second order nature of the filter ensures a more gradual roll-off of high-frequency components, resulting in a stable and continuous output. The PWM output (50$\%$ duty cycle) from Arduino and filtered dc output are shown in figure \ref{fig5}b. The measurements were performed with a 1 GHz Oscilloscope (Yokogawa model no. ). The transfer function H(s) and cutoff frequency $\omega_0$ of the filter is given as,

\begin{equation}
   H(s) = \frac{\omega_o^2}{s^2+2\alpha s+\omega_0^2}
\end{equation}
\begin{equation}
\omega_0 = \frac{1}{C_1C_2R_1R_2}   
\end{equation}
where $s = j\omega$ and $\alpha$ is attenuation constant.

To further verify the dc characteristics of the Sallen-Key output, we measured the output of the outer electrode power supply for a range of control signal amplitudes. The plot in figure \ref{fig6} shows a perfectly linear response of the power supply controlled using the frequency filtered PWM output of the Arduino microcontroller.

As discussed above, digital i/o pins of Arduino microcontroller were employed to record the TTL pulse outputs from the CFD and Charge integrator. Typical count rate from the CEM detector (and CFD) during the experiments was kept below 1 kHz. The projectile beam current was also $\sim$ 50 nA. This results in TTL output frequency of $\sim$ 500 Hz from the charge integrator. The clock frequency of Arduino Uno microcontroller is 16 MHz, which is high enough for faithful recording of these data at the digital i/o pins. Nevertheless, an experimental validation is useful for assigning confidence in the measured electron energy spectrum. A characteristic feature of ion-atom collision process is the probabilistic nature of electron emission. Therefore, the measured electron count, N,  at a fixed energy (CMA voltage) follows Poisson distribution and over multiple measurements the number N has an uncertainty of $\sqrt{N}$, under otherwise identical experimental conditions. We have measured this distribution for various pass energies of the CMA. For the measurement, the outer electrode potential was kept at a fixed voltage, and electron count data (the CFD output) was recorded multiple times for a fixed number of pulses received from the charge integrator module. In figure \ref{fig7}, we have shown one such measured histogram for electrode voltage = -50 V. The histogram is fitted well with a Gaussian distribution and the standard deviation ($\sigma$) is comparable to $\sqrt{N}$, where N is the peak of the Gaussian distribution.   

\begin{figure}
    \centering
    \includegraphics[width = \textwidth]{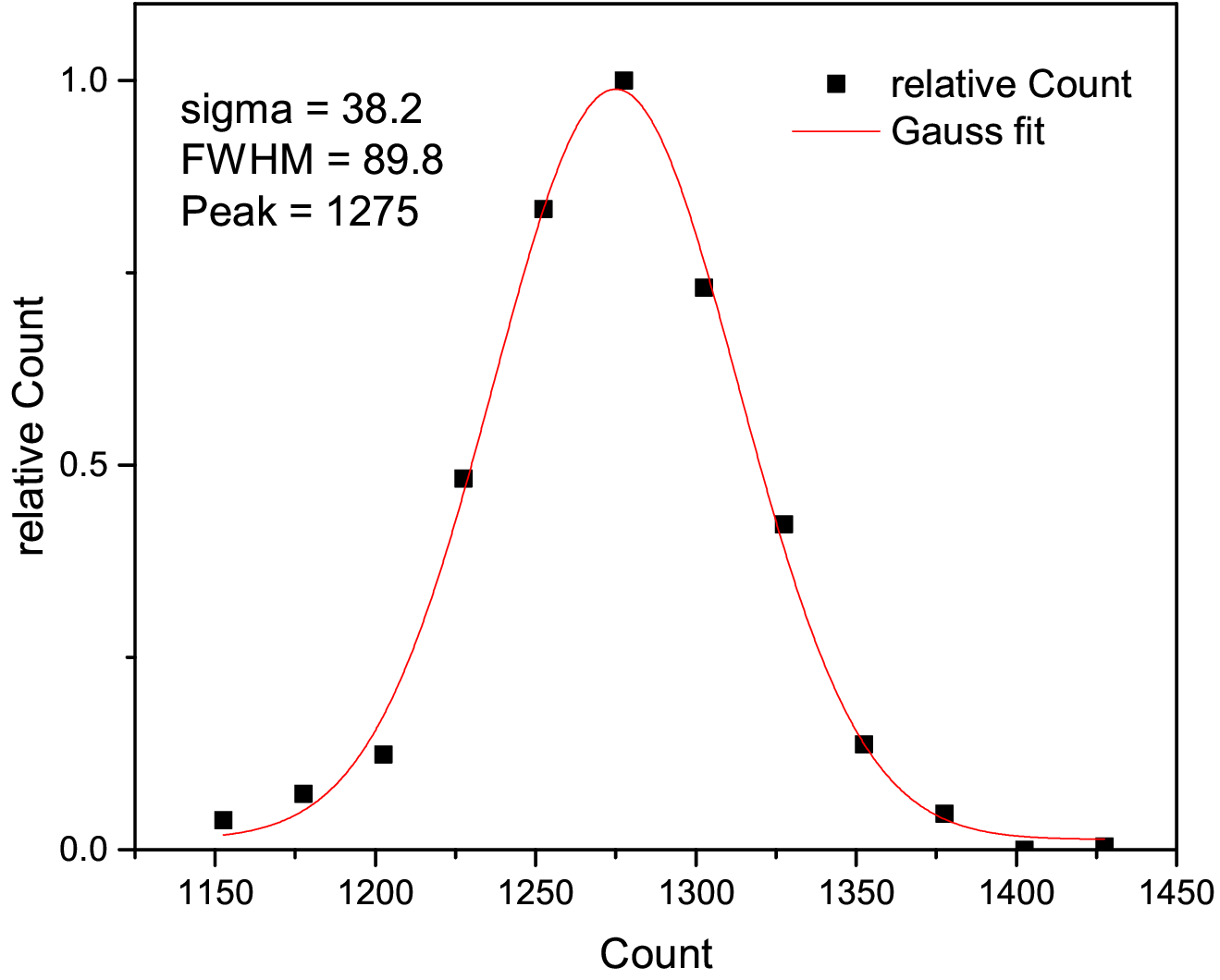}
    \caption{ Electron counts statistics while keeping the outer electrode at -50 V. Measurement has been done 911 times.}
    \label{fig7}
\end{figure}

The process of control signal generation as well as simultaneous recording and storage of the CEM detector and charge integrator outputs was automated using a custom written Python \cite{python} script. To facilitate real-time data communication between the experimental setup and the PC, the pyserial 3.5 \cite{pyserial} library is employed. The Arduino board was programmed to send the necessary information through the serial port, accessible to the PC through the Python script. The Arduino script continuously sends the relevant data, such as electron detection events from the CFD and the beam current from the charge integrator, through the serial port. %On the Python side, a script is implemented to open and monitor the serial port, reading the incoming data stream.The decoded signals representing electron detection events and beam current are then processed in Python. 
The Python script utilizes Matplotlib module in Python to create online electron energy spectrum, therefore enabling visualization of the experimental data in real-time.

\begin{figure}
    \centering
    \includegraphics[width = \textwidth]{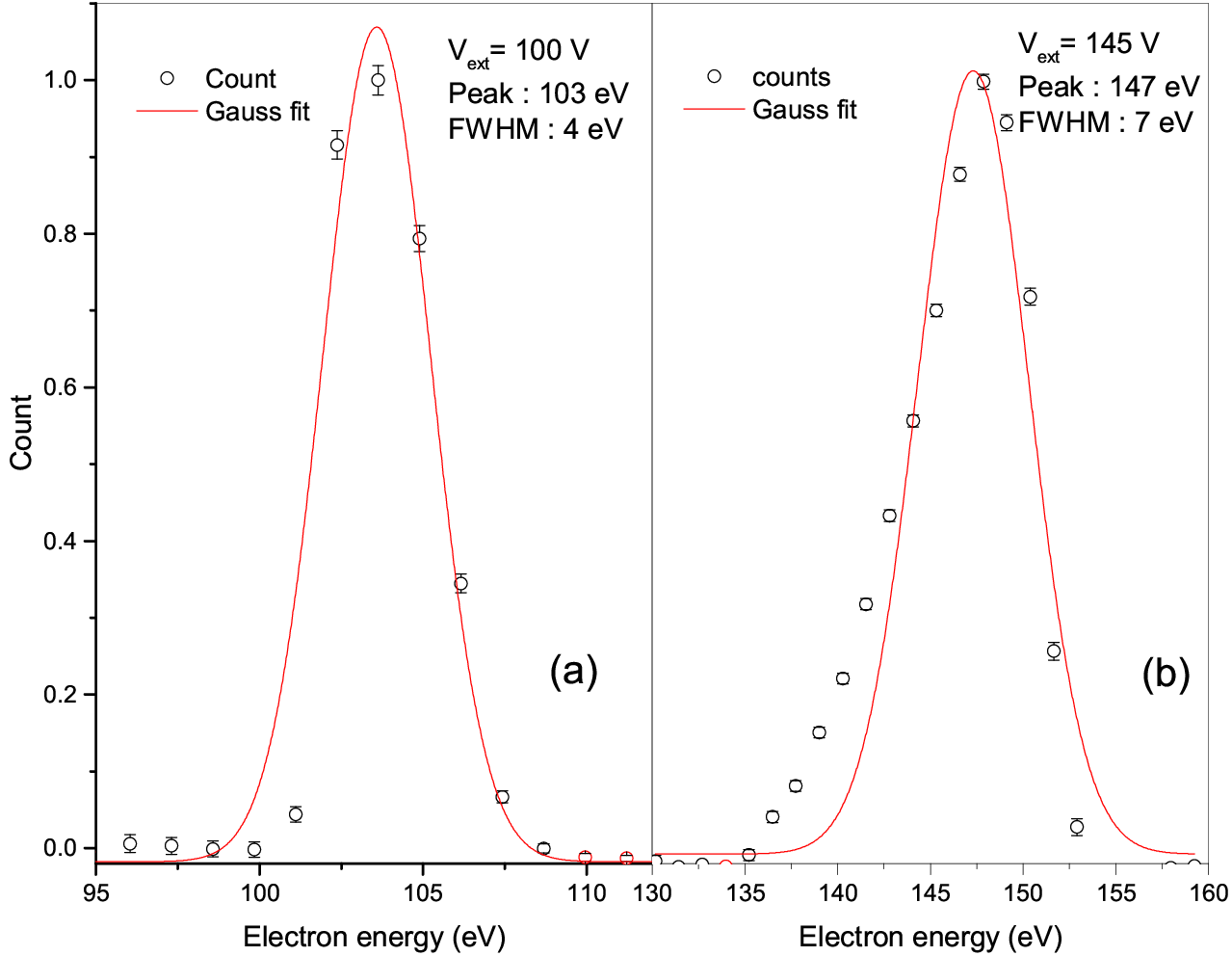}
    \caption{Electron count distribution passing from both slit from an electron gun with extraction voltage kept at (a) 100 V (b) and 145 V. Electron gun filament was at 3.2 V and 1.7 A. }
    \label{fig8}
\end{figure}

\subsection{CMA energy resolution}

For estimating the resolution of the CMA, we used a diverging beam electron gun in a separate experiment. The electron gun was placed along the axis of the CMA and electron spectra were measured for several energies of the emitted electrons. The emitted electron energy had a spread of $\mathrm{\sim 10\%}$. The energy resolution of a CMA spectrometer is given as:

\begin{equation}
   \mathrm{\frac{\Delta E}{E} = \frac{w}{L}} 
\end{equation}

where $\mathrm{w}$ is the slit width and $\mathrm{L}$ is the distance between the slits.

The measured electron energy spectra are sown in figure \ref{fig8} for two electron energies, 100 eV and 145 eV. Using a Gaussian fit to the measured electron spectrum, we have estimated the energy resolution of the CMA $\mathrm{\sim 3 - 4\%}$. A similar resolution was estimated from SIMION simulation.

\section{Calibration and measurements}

\begin{figure}[]
\includegraphics[width=\textwidth]{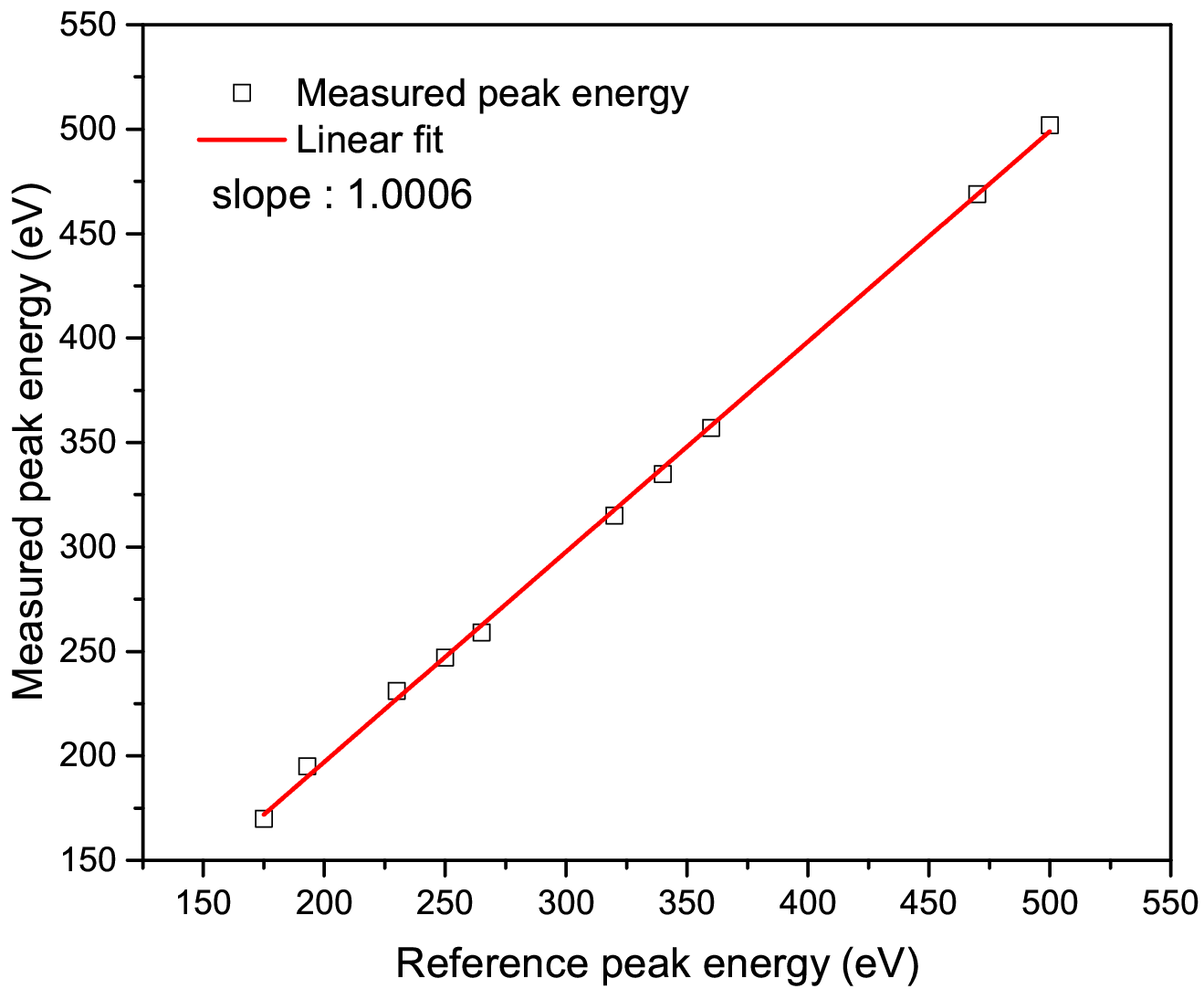}
	\caption{Measured Auger electron energy peak  of Ar, Oxygen in $CO_2$, Carbon in $CH_4$ and nitrogen in N$_2$ on Y axis and peak from reference on x axis by 1 MeV proton bombardment are shown. Straight line fit slope is 1.02 with standard error 0.026. Error bar of 3\% are shown}
	\label{fig9}
\end{figure}

\begin{figure}
    \centering
    \includegraphics[width =0.9\textwidth]{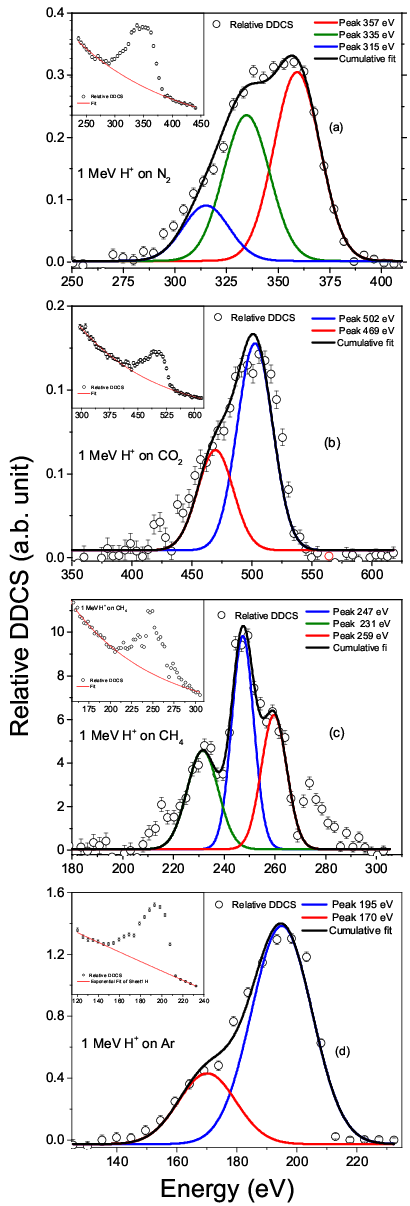}
    \caption{Measured electron energy spectrum of N$_2$ by 1.2 MeV proton bombardment (at Top). And $CO_2$, $CH_4$ and Ar ( from Top to bottom) by 1 MeV proton bombardment.}
    \label{fig10}
\end{figure}

\begin{figure}
    \centering
    \includegraphics[width =\textwidth]{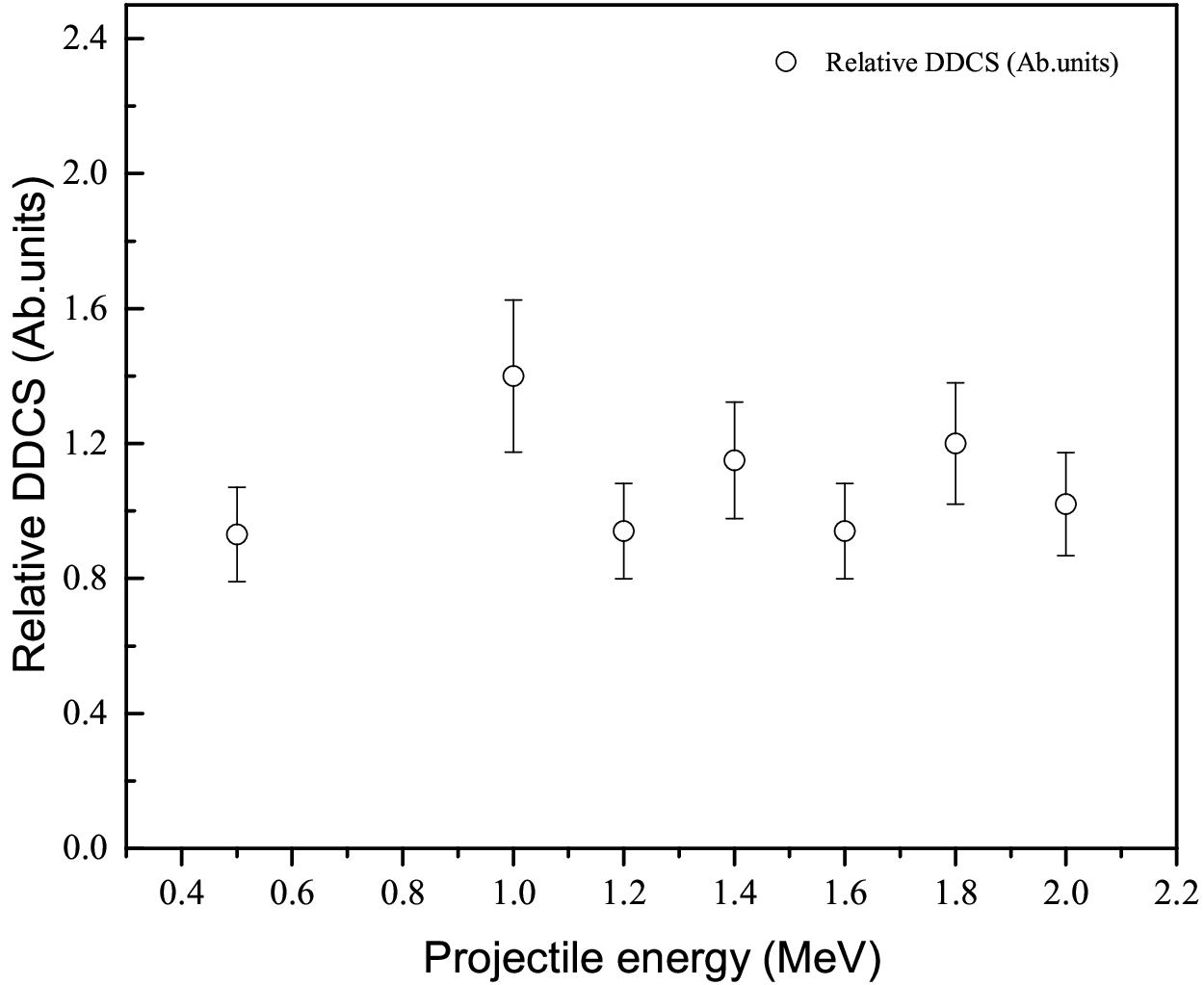}
    \caption{Total cross section after integrating DDCS for different projectile energy.}
    \label{fig11}
\end{figure}

For calibration of the CMA electron spectrometer we measured the Auger spectra for various atomic and molecular targets in collision with 1 MeV proton beam. In figure \ref{fig10} we have shown the electron spectra corresponding to K-LL Auger emission from CH$_4$, N$_2$, CO$_2$ molecules and L-MM Auger emission from Ar atoms. In each spectrum, the electron counts have been normalized with respect to the projectile beam intensity and target gas pressure. The spectrum represents relative differential cross section for electron emission. Geometry of CMA restricts collection of electrons in a narrow range of angle $\mathrm{\theta = 35.5^o}$ with respect to the projectile ion direction, however, there is near complete ($\mathrm{2\pi}$) collection in the azimuthal plane (normal to the spectrometer axis). Therefore, the relative differential cross section is given as:

\begin{equation}
  \frac{d^2\sigma}{dE_ed\theta} = \frac{N_e - N_{nkg}}{N_pP}
\end{equation}

where, $N_e$ is the number of target electrons (with gas) detected at energy $E_e$, $N_{bkg}$ is the number of background electrons (without gas), $E_e$ is the emitted electron energy, $N_P$ is the number of projectile ions and P is the target gas pressure. An ionization gauge was used to measure the gas pressure and hence absolute number density of target gas molecules/atoms could not be measured in the present experiments.

The energy spectra were measured in a limited energy range around the Auger emission energy peak region in all cases. The spectra (see inset in figure \ref{fig10}) show the characteristic features of secondary electron emission following collision with energetic ions \cite{MISRA2009}. The electron emission cross section decreases monotonically as a function of electron kinetic energy. This is attributed to the binary collision process. A broad peak corresponding to the Auger process is visible riding over this continuum background. In order to extract the contribution from Auger peaks, one can approximate the continuum background in the peak region by a polynomial function and subtract it from the measured energy spectrum. The interpolated continuum background is shown as solid curve in the inset of figure \ref{fig10} and the Auger peaks after background subtraction are shown in the main panel. The Auger energy distribution is a broad peak and consists of several Auger lines corresponding to emission from different L and M sub-shells. More than one peak in the Auger spectrum can be identified. However, the resolution of CMA is not high enough to clearly resolve the sub-shell contribution and we have used multiple Gaussian fits to reveal identifiable Auger lines. Additionally, in energetic ion-atom collision the Auger process is also accompanied by multiple ionization of the valance electrons, thereby broadening the Auger region. In table \ref{table1}, we have reported the measured Auger energy values (using the calibration factor from SIMION simulation) are in excellent agreement with the Auger values reported in literature \cite{Kobayashi1976} validating the calibrating factor obtained using the SIMION simulation.

We further studied the projectile energy dependence of $N_2$ K-LL Auger emission yield. The proton beam energy was varied from 0.5 MeV to 2 MeV. Several groups have reported detailed cross section measurements on K-LL auger emission in the later part of last century \cite{Stolterfoht1975,Kobayashi1976}. For low Z elements, the K-LL Auger cross sections are equivalent to the inner shell ionization cross section due to extremely low fluorescence yield leading to x-ray emission. %$\textsl{Ab initio}$ models based on Born approximation \cite{} and Binary encounter approximation \cite{} have been compared with the measured data and a good agreement has been found for proton projectile. 
The K-LL Auger emission cross sections reach a maximum in the present projectile energy range and a variation within 10$\%$ has been reported \cite{Kobayashi1976}. In figure \ref{fig11}, we have shown the integrated yield (relative cross section) for N$_2$ K-LL Auger emission in ionizing collisions with 0.5 MeV to 2 MeV proton beam. The measured relative cross sections do not vary much with in the experimental uncertainty ($\sim$ 15$\%$). This is in agreement with the reported measurements in this projectile energy range for H$^+$ ions. The uncertainty in the measured cross sections in attributed mainly to the fluctuation in the target gas pressure.

%We have performed measurement on Ar, N$_2$, CO$_2$ and CH$_4$ gas target to test the working of CMA. Electron spectrum for a 1 MeV proton beam are shown in fig. In fig a Argon spectrum is shown. Due to limited resolution we can not separate multiple peak in the spectrum .

%TO test the performance of the set-up we have measured count statistics for a fixed pass energy which is shown in figure. For good performance of the spectrometer the FWFM of the distribution shuld be close or less to the twice of the square root of the peak counts. and in fig b we have shown the CMA elect rod bias power supply output vs the control signal which was coming from the Sallen key-filter. It is linear which confirms the stability for the electrode voltage.%

%\begin{figure}
 %   \centering
  %  \includegraphics[width =\textwidth]{PEN2Peaks.eps}
   % \caption{Electron energy distribution of N$_2$ after background subtraction for different projectile energy }
    %\label{PEN2Peaks}
%\end{figure}
\begin{table}
\caption{Auger peak energy for different targets. The measured values have been compared with those give in reference \cite{Kobayashi1976}}\label{table1}
\begin{ruledtabular}
\begin{tabular}{ p{1.5cm} p{1.5cm} p{1.5cm} p{1.5cm}  }
\textrm{proton energy (MeV)}&
\textrm{Target gas}&
\textrm{Auger $e^-$ energy (eV)}&
\textrm{Reference \cite{Kobayashi1976}}\\
\colrule
1 & $N2$ & 357,335,315 & 360,340,320\\
1 & $CO_2$ & 502,469 (O,KLL)& 500,470\\
1 & $CH_4$ & 247,231,259 (C, KLL) & 250,230,265\\
1 & $Ar$ & 195,170 (LMM) & 193,175\\
\end{tabular}
\end{ruledtabular}
\end{table}

%\begin{table}%The best place to locate the table environment is directly after its first reference in text
%\caption{\label{tab:table1}%
%}

%\begin{ruledtabular}
%\begin{tabular}{ p{1.5cm} p{1.5cm} p{1.5cm}}
%\textrm{H$^+$ energy (MeV)}&
%\textrm{$\sigma_{KLL}\times$ 10$^{-19}$ cm$^2$}&
%\textrm{$\sigma_{KLL}\times$ 10$^{-19}$ cm$^2$}\\
%\colrule
%0.5 & 3.9 & --\\
%1 &  5.9 & 5.2$\pm$ 1.0\\
%1.2 & 3.9 & --\\
%1.4 &  4.8 & 5.5$\pm$ 1.1\\
%1.6 &  3.9 & --\\
%1.8 &  5.0 & 6.0 $\pm$ 1.2\\
%2.0 &  4.3 & --\\
%\end{tabular}
%\end{ruledtabular}
%\end{table}

\section{conclusion}
We have presented the design, construction, and testing of a cylindrical mirror analyzer electron spectrometer and associated data acquisition system. The spectrometer is capable of measuring electron spectrum with an energy  resolution of $\sim$ 4$\%$ and nearly 2$\pi$ collection of electrons in the azimuthal plane. The spectrometer performance was evaluated by measuring electron spectra, following collision with MeV energy proton beam, for $\mathrm{N_2, CO_2, CH_4 and Ar}$ in the K-LL (L-MM for Ar) Auger energy range. Energy of Auger peaks measured for different targets match well with the reported values in literature.

\section{Acknowledgments}
\begin{acknowledgments}
The Authors would like to thank Sahan Sykam and Sandeep Bari for smooth operation of the tandetron accelerator facility during the experiments. AHK would also like to acknowledge support from DST-SERB grant no. ECR/2017/002055.
\end{acknowledgments}

%\nocite{*}
\bibliographystyle{utphys}
\bibliography{output}% Produces the bibliography via BibTeX.
\end{document}